\documentclass[aps, pre, showpacs, showkeys, twocolumn]{revtex4-1}

\usepackage{amsmath, amssymb, amsthm, amsfonts, latexsym}
\usepackage{bm, amsfonts, mathtools}
\usepackage{graphicx, color, wasysym}
\usepackage{textcomp}
\usepackage{epsfig}
\usepackage{appendix}

\def\ulamek#1#2{\mbox{\normalfont$\frac{#1}{#2}$}}

\DeclareMathOperator{\D}{d\!}
\DeclareMathOperator{\E}{e}

\begin{document}

\title[Relativistic Heat Equation via L\'{e}vy stable distributions: Exact Solutions]{Relativistic Heat Equation via L\'{e}vy stable distributions: Exact Solutions}

\author{K.~A.~Penson}
\email{penson@lptl.jussieu.fr}
\affiliation{Sorbonne Universit\'{e}s, Universit\'{e} Pierre et Marie Curie (Paris VI), CNRS UMR 7600, Laboratoire de Physique Th\'{e}orique de la Mati\`{e}re Condens\'{e}e (LPTMC), Tour 13-5i\`{e}me \'{e}t., B.C. 121, 4 pl. Jussieu, F 75252 Paris Cedex 05, France}

\author{K.~G\'{o}rska}
\email{katarzyna.gorska@ifj.edu.pl}
\affiliation{H. Niewodnicza\'{n}ski Institute of Nuclear Physics, Polish Academy of Sciences, Division of Theoretical Physics, ul. Eliasza-Radzikowskiego 152, PL 31-342 Krak\'{o}w, Poland}

\author{A.~Horzela}
\email{andrzej.horzela@ifj.edu.pl}
\affiliation{H. Niewodnicza\'{n}ski Institute of Nuclear Physics, Polish Academy of Sciences, Division of Theoretical Physics, ul. Eliasza-Radzikowskiego 152, PL 31-342 Krak\'{o}w, Poland}

\author{G. Dattoli}
\email{giuseppe.dattoli@enea.it}
\affiliation{ENEA - Centro Ricerche Frascati, via E. Fermi, 45, IT 00044 Frascati (Roma), Italy}


\begin{abstract}
We introduce and study  an extension of the heat equation relevant to relativistic energy formula involving square root of differential operators. We furnish exact solutions of corresponding Cauchy (initial) problem using the operator formalism invoking one-sided L\'{e}vy stable distributions. We note a natural appearance of Bessel polynomials which allow one the obtention of closed form solutions for a number of initial conditions. The resulting relativistic diffusion is slower than the non-relativistic one, although it still can be termed a normal one. Its detailed statistical characterization is presented in terms of exact evaluation of arbitrary moments and is compared with the non-relativistic case. 
\end{abstract}



\maketitle

The ordinary diffusion problem is customarily treated in terms of parabolic differential equation, linking first order derivative in time and second order derivative in space of a distribution function, which in the case of the heat equation, represents a physical quantity like the behavior of temperature. The extension of the equation to include relativistic effects has been the topic of intense debate in the past, see \cite{CJ, IgnOS, JoPr, GJKS}, and particularly \cite{DH} for extensive bibliography. The authors of \cite{GJKS} have developed and interesting point of view displaying how the Brownian motion underlying ordinary diffusion is extended to relativistic realm. According to \cite{GJKS} the problem is ruled by the telegrapher equation, thus finding agreement with the original Cattaneo's treatment of the problem \cite{Catt}. The interest of their analysis stems from the fact that they emphasize the intimate connection between the stochastic process described by the telegrapher equation and the Dirac equation.

In this note we develop an alternative point of view which takes advantage of the concepts of anomalous diffusion \cite{MeSo} governed by the evolution equations involving square root pseudo-differential operators.  
We postulate and study in this note the following one-dimensional differential equation for the distribution function $\psi(\xi, \tau)$ which we believe would be appropriate in the relativistic (R) situation:
\begin{equation}\label{30/07-1}
\lambda\partial_{\tau} \psi(\xi, \tau) = m_{0}c^{2}\{1\! -\! [1\! - \!(\!\ulamek{\lambda}{m_{0}c}\!)^{2}\partial_{\xi}^{2}]^{1/2}\} \psi(\xi, \tau), 
\end{equation}
with $\xi$ and $\tau$ respectively space and time variables, and where $m_{0}$ is the rest mass and $\lambda > 0$ has the dimension of action. The conventional diffusion equation \cite{DWidder}
\begin{equation}\label{21/09-2}
\partial_{\tau} \psi(\xi, \tau) = D \partial_{\xi}^{2} \psi(\xi, \tau)
\end{equation}
can be obtained from the non-relativistic (NR) Schr\"{o}dinger equation by means of the Wick rotation $\tau\to -i t$. We observe that Eq. \eqref{30/07-1} can be analogously produced by performing the Wick rotation on the R Schr\"{o}dinger (i.e. the Salpeter \cite{KR}) equation with substracted rest energy $m_{0}c^{2}$. The NR limit of Eq. \eqref{30/07-1}, i.e. the $1/c$ expansion in the spirit of \S 33 of \cite{BLP}, results in Eq. \eqref{21/09-2} with the identification $D=\lambda/(2m_{0})$. Subsequently we introduce in Eq. \eqref{30/07-1} the dimensionless variables $t = m_{0}c^{2}\tau/\lambda$ and $x = m_{0}c\xi/\lambda$ and we end up with
\begin{equation}\label{30/07-2}
\partial_{t} \psi(x, t) = (1-\sqrt{1-\partial^{2}_{x}}) \psi(x, t),
\end{equation}
which is formally of infinite order in $\partial^{2}_{x}$. For the purpose of this work we shall refer to Eq. \eqref{30/07-2} as a relativistic heat equation. In Quantum Mechanics the difficulty to treat the square root in Eqs. \eqref{30/07-1} and \eqref{30/07-2} (so-called pseudo-differential operators \cite{Lamm, Bab}) is usually resolved by introducing the wave functions with several components, like in Pauli and in Dirac equations \cite{Lamm, Bab}. However the interest for equations involving pseudo-differential operators arose even before the inception of the Dirac equation. It can be traced back to H. Weyl \cite{Weyl}. The fact that Eqs. \eqref{30/07-1} and \eqref{30/07-2} correctly reproduce the NR limit is reminiscent of the R extension of the analogy between the Schr\"{o}dinger and the heat equations, carried out in 1984 in \cite{GJKS}, by identifying the underlying Poisson processes. Their real version obeys the telegrapher's equation, which, when analytically continued, produce the Dirac equation. To see how this circumstance mirrors in our context, we square the operator of Eq. \eqref{30/07-2} and we get: 
\begin{equation}\label{14/09-1}
(\partial_{t}^{2} + \partial_{x}^{2} - 2 \partial_{t}) \tilde{\psi}(x, t) = 0,
\end{equation}
where the ensemble of solutions of $\tilde{\psi}(x, t)$ contains the solutions of Eq. \eqref{30/07-2}. Note that Eq. \eqref{14/09-1} is similar but not identical to the telegrapher's equation \cite{JoPr, Pov}, also referred to as the Cattaneo equation \cite{Catt}. We can shed another light on Eqs. \eqref{30/07-1} and \eqref{30/07-2} by modifying the definition of constants in Eq. \eqref{30/07-1}: introduce $\theta$ and $\Delta$ with dimensions of time and length respectively. Then $\theta\partial_{t}F(x, t) = [1-\sqrt{1-\Delta^{2} \partial_{x}^{2}}]F(x, t) \approx (\Delta^{2} \partial_{x}^{2}/2)F(x, t)$, leading to $D=\Delta^{2}/(2\theta)$. If $\Delta = v \theta$ then it is a kind of free path in the characteristic time $\theta$. The expansion in $\Delta$ means that $\Delta$ is smaller than the typical diffusion lenght. (With the above notation Eq. (9) in \cite{GJKS} becomes $\theta \partial_{t} h(x, t) = [(1+\Delta^{2}\partial_{x}^{2})^{1/2} - 1]h(x, t)$ which however cannot be treated by our method of Eqs. \eqref{30/07-3}-\eqref{30/07-7}.) Therefore our Eq. \eqref{30/07-2} provides an alternative to the procedure of \cite{GJKS}.

Here, we keep the scalar nature of $\psi(x, t)$ in Eq. \eqref{30/07-2} and set out to solve it exactly. Since Eq. \eqref{30/07-2} is linear in $\partial_{t}$ its full solution is obtained by the action of the evolution operator $\hat{U}(t)$ on the initial condition (i.c.) $\psi(x, 0) \equiv f(x)$:
\begin{equation}\label{30/07-3}
\psi(x, t) = \hat{U}(t) f(x) \equiv \exp[t(1-\sqrt{1-\partial^{2}_{x}})] f(x).
\end{equation}
The objective of the present work is to elucidate the properties of $\hat{U}(t)$ and to derive a number of exact solutions of Eq. \eqref{30/07-2}. To this end we make contact with the exponential generating function of Bessel polynomials $B_{n}(t)$~\cite{Grosswald}:
\begin{equation}\label{30/07-4}
\sum_{n=0}^{\infty} \frac{z^{n}}{n!} B_{n}(t) = \exp[t(1 - \sqrt{1 - 2z})].
\end{equation}
Then we define the L\'{e}vy-Smirnov function \cite{Uchaikin} $g_{1/2}(\xi) = \exp[-1/(4\xi)]/(2\sqrt{\pi} \xi^{3/2})$, $\xi\geq 0$, the canonical case of one-sided L\'{e}vy stable distributions \cite{PG}, whose Laplace transform is known \cite{Doetsch}:
\begin{equation}\label{30/07-5}
\E^{-\eta\sqrt{p}} = \int_{0}^{\infty} g_{1/2}(\xi) \E^{-\eta^{2} p \xi} \D \xi, 
\end{equation}
for ${\rm Re}(p) > 0$ and $\eta > 0$. Combining Eqs. \eqref{30/07-4} and \eqref{30/07-5} we obtain the relation between $B_{n}(t)$ and $g_{1/2}(\xi)$:
\begin{equation}\label{30/07-6}
B_{n}(t) = (2 t^{2})^{n} \E^{t} \int_{0}^{\infty} g_{1/2}(\xi) \E^{-\xi t^{2}} \xi^{n} \D\xi.
\end{equation}
Using now Eq. \eqref{30/07-4} for $z = \partial^{2}_{x}/2$ the following form of Eq. \eqref{30/07-3} emerges
\begin{equation}\label{30/07-7}
\psi(x, t) = \left[\sum_{n=0}^{\infty} \frac{B_{n}(t)}{2^{n} n!} \partial_{x}^{2n}\right] f(x).
\end{equation}
For many i. c. $f(x)$ the series of Eq. \eqref{30/07-7} can be summed explicitly, as the Bessel polynomials (in Carlitz form) are known: $B_{n}(t) = \sum_{k=1}^{n} \frac{(2n-k-1)!}{(k-1)! (n-k)!} \frac{t^{k}}{2^{n-k}}$.
If $f(x)$ is defined for $|x| < \infty$ then the solution of the NR heat equation, Eq. \eqref{21/09-2} for $D=1/2$, is given by 
\begin{equation}\label{30/08-1}
\psi_{\rm NR}(x, t) = \E^{\frac{t}{2}\partial_{x}^{2}} f(x) = \int_{-\infty}^{\infty} \frac{\E^{-\ulamek{(x-u)^{2}}{2t}}}{\sqrt{2\pi t}} f(u) \D u, 
\end{equation}
$t > 0$, which is the Gauss-Weierstrass (GW) transform \cite{DWidder} of $f(x)$. In contrast, the R case, for \textit{the same} initial condition $f(x)$, according to Eqs. \eqref{30/07-5} and \eqref{30/07-6}, is given by
\begin{align}\label{30/07-9}
\psi_{\rm R}(x, t) & = \E^{t} \int_{0}^{\infty} g_{1/2}(y) \E^{-y t^{2}} \E^{y t^{2} \partial_{x}^{2}}f(x) \D y \nonumber\\
& = \E^{t} \int_{0}^{\infty} g_{1/2}(y) \E^{-y t^{2}} \psi_{\rm NR}(x, 2y t^{2}) \D y,
\end{align}
i.e. it is expressed through an integral operator involving the $g_{1/2}(y)$ function, thus linking the R and NR solutions. We shall compare now the NR and R equations for i.c. $f^{(1)}(x) = \E^{-x^{2}}/\sqrt{\pi}$, i.e. the normalized Gaussian. (The superscripts in $f^{(j)}(x)$ enumerate the i.c.'s considered in this work.) As it is well known the NR time-evolved Gaussian is again a Gaussian, as seen from the Glaisher formula after $t/2$: $\psi_{\rm NR}^{(1)}(x, t) = \frac{1}{\sqrt{\pi}} \frac{1}{\sqrt{1+2t}} \exp(-\frac{x^{2}}{1+2t})$ \cite{Dat}. By the same token, the corresponding R exact solution of Eq. \eqref{30/07-9} is given by the following integral:
\begin{equation}\label{30/07-10}
\psi_{\rm R}^{(1)}(x, t) = \int_{0}^{\infty}\!\! g_{1/2}(y) \E^{t-y t^{2}} \frac{\exp(-\frac{x^{2}}{1+4t^{2} y})}{\sqrt{1 + 4t^{2} y}} \frac{\D y}{\sqrt{\pi}},
\end{equation}
which is a kind of integral convolution with $g_{1/2}(y)$. 
\begin{figure}[!h]
\includegraphics[scale=0.36]{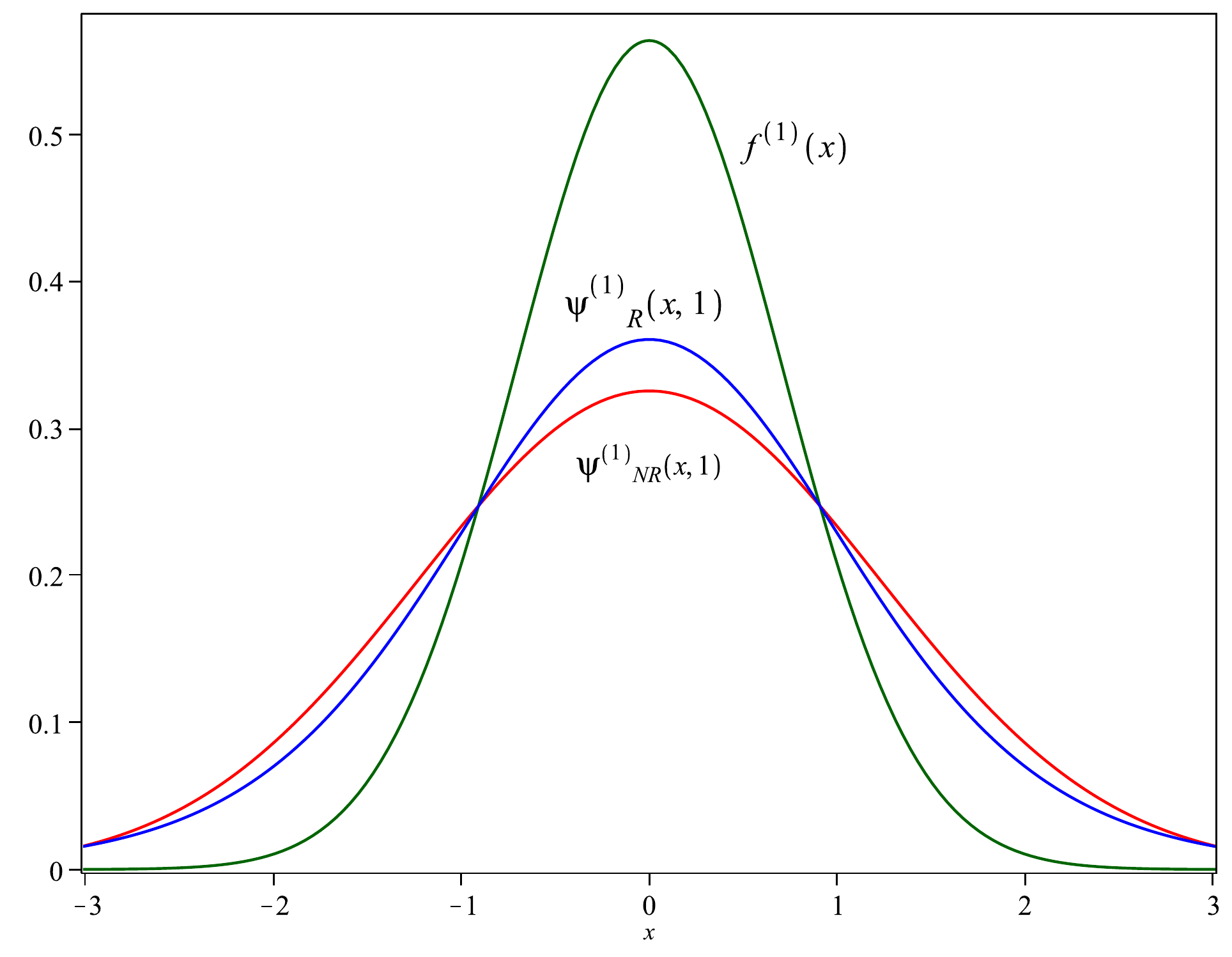}
\caption{\label{fig1} (Color online) Comparison of R and NR distributions with i.c. $f^{(1)}(x)$ for $t=1$: $\psi^{(1)}_{\rm R}$ has less evolved than $\psi^{(1)}_{\rm NR}$. This fact holds for all $t$.}
\end{figure}
In Fig. \ref{fig1} we compare $f^{(1)}(x)$, $\psi_{\rm NR}^{(1)}(x, 1)$ and $\psi_{\rm R}^{(1)}(x, 1)$ from Eq. \eqref{30/07-10}. We observe that the NR evolution is more rapid than the R one: for all times the R curves are less flattened that the corresponding NR ones. Similar results hold for i.c. which are not Gaussians, s. Fig. \ref{fig5} below. Additional informations about the so obtained distributions are contained in the averages of type $\langle x^{2p}\rangle_{\rm R, NR}(t) = \int_{-\infty}^{\infty} x^{2p} \psi_{\rm R, NR}(x, t) \D x$, $p=1, 2, \ldots$. For NR case the calculations of $\langle x^{2p}\rangle_{NR}(t)$ are standard and for a general i.c. $f(x)$ they yield $\langle x^{2}\rangle_{\rm NR}(t) = \int_{-\infty}^{\infty} u^{2} f(u) du + t \int_{-\infty}^{\infty} f(u) du$, implying that both integrals $\int_{-\infty}^{\infty} u^{M} f(u) du$, $M=0, 2$ should be converging if the description in terms of $\langle x^{2}\rangle$ is adopted. In these cases $\langle x^{2}\rangle_{\rm R}(t) = \int_{-\infty}^{\infty} x^{2}\psi_{\rm R}(x, t) dx \equiv \langle x^{2} \rangle_{\rm NR}(t)$, as can be seen from Eq. \eqref{30/07-9} by elementary integration. It means that the R diffusion is normal. Below we shall also display explicit solutions in which $\langle x^{2}\rangle_{\rm R}(t)$ is diverging for all $t$. According to Eq. \eqref{30/07-9} the R averages read: 
\begin{equation}\label{30/07-11}
\langle x^{2p}\rangle_{\rm R}(t) = \E^{t} \int_{0}^{\infty} g_{1/2}(y) \E^{-y t^{2}} \langle x^{2p} \rangle_{\rm NR}(2yt^{2}) \D y.
\end{equation}
For $p=1$ and $f^{(1)}(x)$, Eq. \eqref{30/07-11} furnishes $\langle x^{2}\rangle_{\rm R}^{(1)}(t) = \langle x^{2}\rangle_{\rm NR}^{(1)}(t) = 1/2 + t$, indicating that both NR and R diffusions are of normal type. More refined characteristics of the R diffusion can be  deducted from higher-order averages for $p > 1$, like kurtosis $\kappa(t) = \langle x^{4}\rangle(t)/[\langle x^{2}\rangle(t)]^{2} - 3$, see \cite{MeSo, Wes}. It can be evaluated using $\langle x^{4}\rangle_{\rm R}(t) = \langle x^{4}\rangle_{\rm NR}(t) + 3t$ which leads, for general i. c. $f(x)$, to $\kappa_{\rm R}(t) = \kappa_{\rm NR}(t) + 3t/(\int_{-\infty}^{\infty} x^{2} f(x) \D x + t)^{2}$. This last result implies that for arbitrary, but the same i. c. $f(x)$, $\kappa_{\rm R}(t) \geq \kappa_{\rm NR}(t)$.

\begin{figure}[!h]
\includegraphics[scale=0.42]{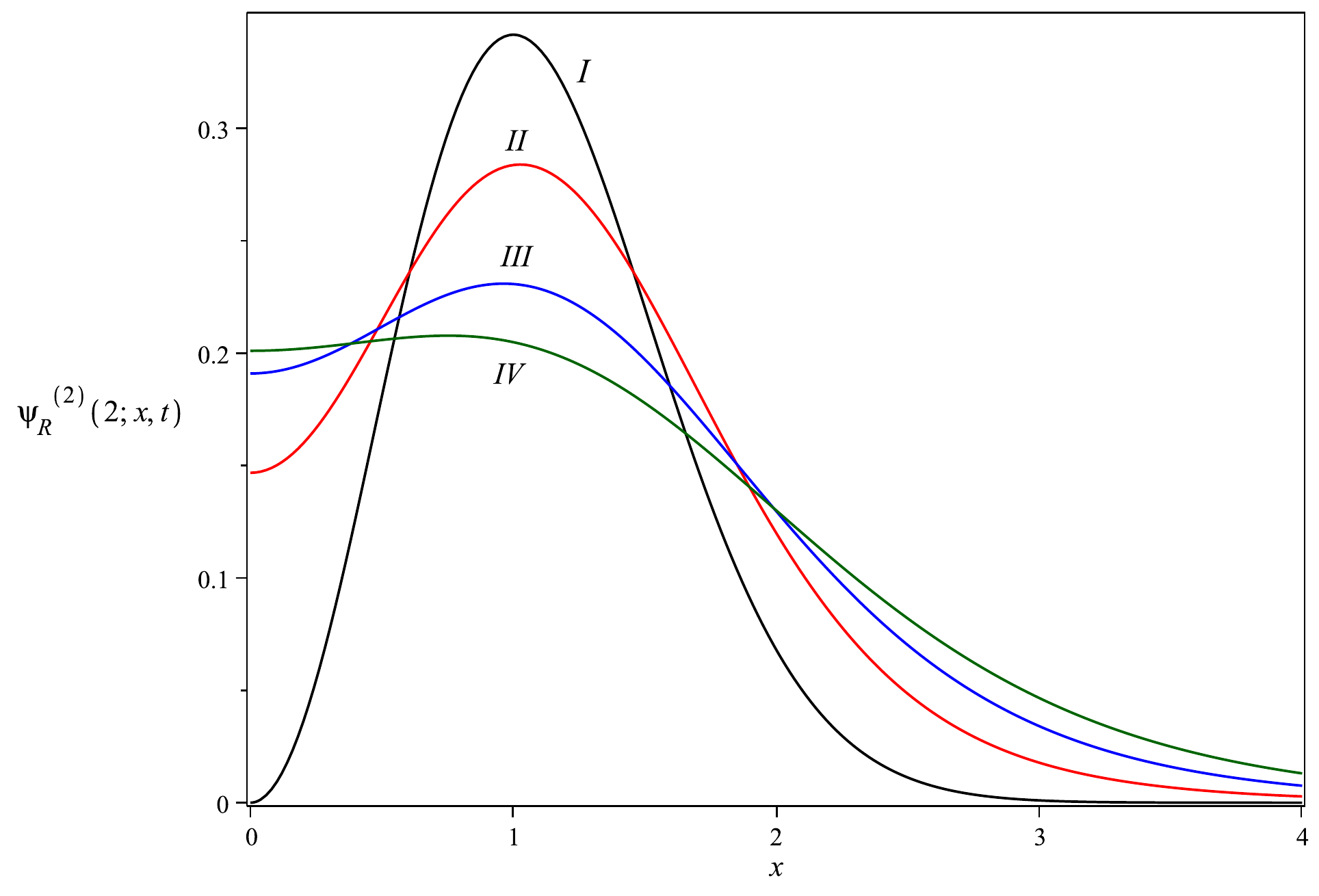}
\caption{\label{fig2} (Color online) Plot of $\psi_{\rm R}^{(2)}(2; x, t)$ given by Eq. \eqref{30/07-13} for $t=0$ (black line, I), $t=0.5$ (red line, II), $t=1$ (blue line, III), and $t=1.5$ (green line, IV).}
\end{figure}

We proceed by constructing exact solutions of Eq. \eqref{30/07-7} for a number of i. c. other than the Gaussian. The first instance is $f^{(2)}(r; x) = 2 x^{r}\exp(-x^{2})/\Gamma(\frac{r+1}{2})$, $r=1, 2, \ldots$, $0 \leq x < \infty$. As an intermediate step we first treat the unnormalized i.c. $h^{(2)}(p; x) = H_{p}(x) \exp(-x^{2})$, $0 \leq x < \infty$, $p=1, 2, \ldots$ where $H_{p}(x)$ are the conventional Hermite polynomials \cite{NIST}. We use now Eq. \eqref{30/07-7}, and observing that $\partial_{x}^{2n}[H_{p}(x)\exp(-x^{2})] = H_{p + 2n}(x) \exp(-x^{2})$ (s. Eq. (1.20.1.11) of \cite{YuABrychkov}) and employing Eq. \eqref{30/07-6} we obtain the R evolution of $h^{(2)}(p; x)$:
\begin{align} \label{30/07-12}
& \varPhi(p; x, t) = \E^{t-x^{2}}\!\!\int_{0}^{\infty}\!\! g_{1/2}(y) \E^{-y t^{2}}\!\! \left[\sum_{n=0}^{\infty} \frac{(y t^{2})^{n}}{n!} H_{2n+p}(x)\right]\!\!\D y \nonumber\\
& = \E^{t}\int_{0}^{\infty}\!\! g_{1/2}(y)\!\! \left[\frac{\exp(-y t^{2}-\frac{x^{2}}{1+4t^{2}y})}{(1+ 4t^{2} y)^{(p+1)/2}} H_{p}\left(\!\!\ulamek{x}{\sqrt{1+4t^{2} y}}\!\right)\!\right]\!\!\D y,
\end{align}
where in deriving Eq. \eqref{30/07-12} we used Eq. (5.12.1.4) of \cite{APPrudnikov-v2}. Eq. \eqref{30/07-12} permits one to deduce the exact normalized solution $\psi_{\rm R}^{(2)}(r; x, t)$ with i.c. $f^{(2)}(r; x)$ by expressing $x^{r}$ as a finite sum of $H_{p}(x)$'s, compare Eq. (18.18.20) of \cite{NIST}, yielding
\begin{equation}\label{30/07-13}
\psi_{\rm R}^{(2)}(r; x, t) = \frac{r!}{2^{r} \Gamma(\frac{1+r}{2})} \sum_{k=0}^{\lfloor r/2\rfloor} \frac{\varPhi(r-2k; x, t)}{k!(r-2k)!}.
\end{equation}
The time evolution of Eq. \eqref{30/07-13} is exemplified for $r=2$ and for several values of $t$ on Fig. \ref{fig2}. 
We observe a curious behavior at $x=0$: the time evolution lifts up the values of solutions at $x=0$: $\psi^{(2)}_{\rm R}(r; 0, t) > 0$ for all $t > 0$. 

\begin{figure}[!h]
\includegraphics[scale=0.42]{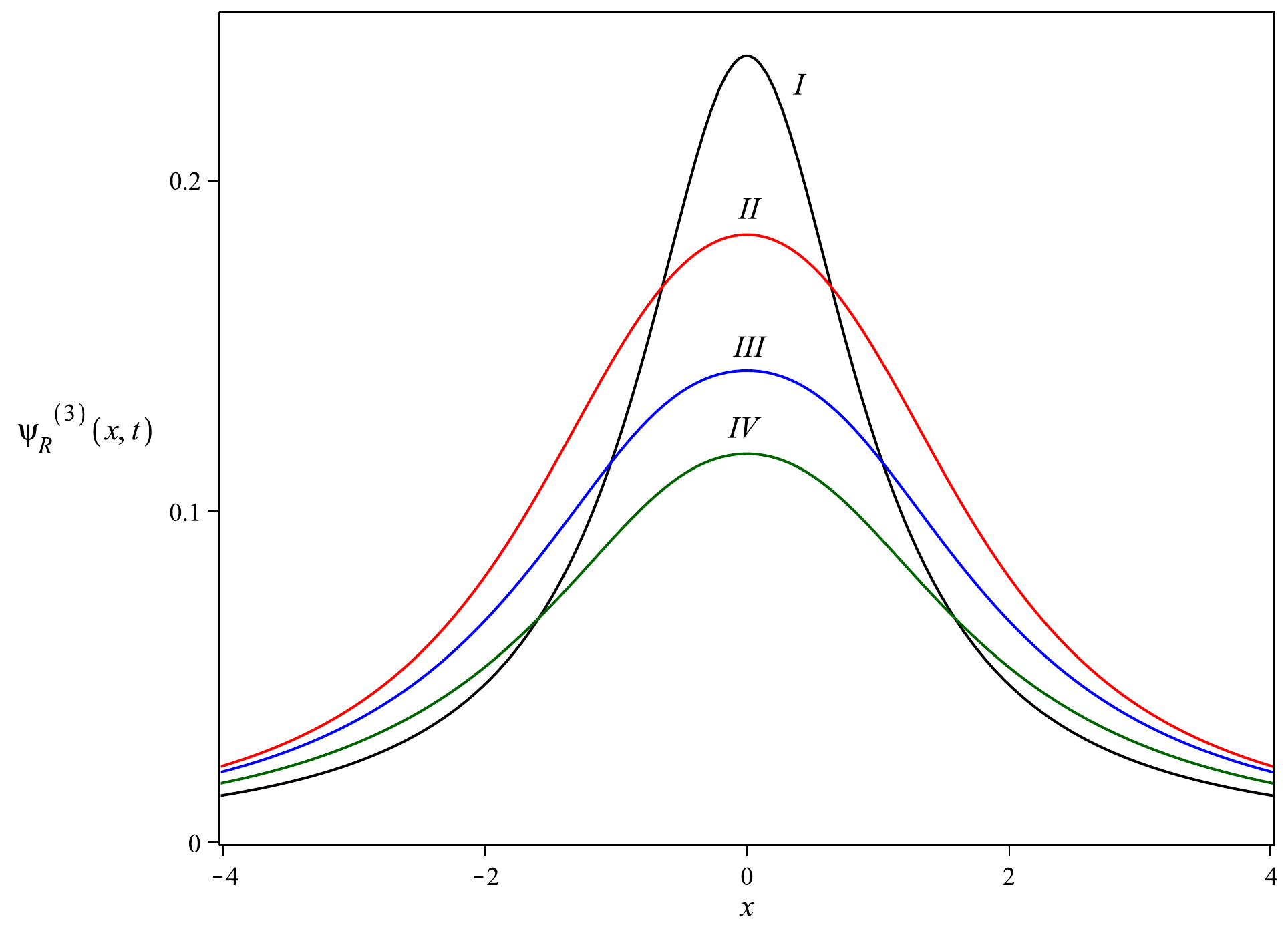}
\caption{\label{fig3} (Color online) Plot of $\psi_{\rm R}^{(3)}(x, t)$ given by Eq. \eqref{30/07-16} for $t=0$ (black line, I), $t=2$ (red line, II), $t=4$ (blue line, III), and $t=6$ (green line, IV).}
\end{figure}

The choice of i.c. $f^{(3)}(x) = [\pi(1+x^{2})]^{-1}$, $|x| < \infty$, (Cauchy function) also allows for exact solution. However here $\int_{-\infty}^{\infty} x^{2} f^{(3)}(x) dx$ is diverging. Going back to Eq. \eqref{30/07-7} we first observe that $\partial_{x}^{2n}[(1+x^{2})^{-1}] = (2n)! (1+x^{2})^{-(n+1)} U_{2n}(-x/\sqrt{1+x^{2}})$, where $U_{k}(z)$, $|z|~<~1$, are the Chebyshev polynomials \cite{GPD}, and the solution can be recast in the form
\begin{align}\label{30/07-14}
\psi_{\rm R}^{(3)}(x, t) &= \frac{\E^{t}}{\pi^{3/2}(1+x^{2})}\! \int_{0}^{\infty}\!\! g_{1/2}(y) \E^{-y t^{2}}\!\! \left[\sum_{n=0}^{\infty} \left(\frac{4 t^{2} y}{1+x^{2}}\right)^{n} \right. \nonumber \\
&\times \left.\Gamma(n+\ulamek{1}{2}) U_{2n}\left(\frac{-x}{\sqrt{1+x^{2}}}\right)\right] \D y.
\end{align}
The inner summation in Eq. \eqref{30/07-14} (in obvious notation), i.e. $r^{(3)}(\tilde{t}, \tilde{x})~\equiv~\sum_{n=0}^{\infty} \tilde{t}^{n} \Gamma(n+\frac{1}{2}) U_{2n}(\tilde{x})$, can be repre-\\sented  in integral form using $U_{2n}(\tilde{x})=[(2n)!]^{-1}\int_{0}^{\infty} \E^{-u} \\\times u^{n} H_{2n}(\tilde{x}\sqrt{u}) \D u$ \cite{BDF} and the previously quoted Eq. (5.12.1.4) of \cite{APPrudnikov-v2}. The result is, for $|\tilde{x}| \leq 1$:
\begin{equation}\label{30/07-15}
r^{(3)}(\tilde{t}, \tilde{x}) = \sqrt{\pi} \int_{0}^{\infty} \frac{\exp[\frac{-u - u^{2} \tilde{t}(1 - \tilde{x}^{2})}{1 + \tilde{t}u}]}{\sqrt{1 + \tilde{t}u}} \D u,
\end{equation}
furnishing, with the auxiliary function $R^{(3)}(t, x, y) \equiv r^{(3)}(\frac{4 y t^{2}}{1+x^{2}}, \frac{-x}{\sqrt{1+x^{2}}})$, ($t\geq 0$, $|x| < \infty$), the final form:
\begin{equation}\label{30/07-16}
\psi_{\rm R}^{(3)}(x, t) = \int_{0}^{\infty}\!\! g_{1/2}(y) \E^{-y t^{2}} \frac{\E^{t} R^{(3)}(t, x, y)}{\pi^{3/2}(1+x^{2})}\D y.
\end{equation}
The time evolution of Eq. \eqref{30/07-16} is illustrated for several values of $t$ on Fig. \ref{fig3}.

\begin{figure}[!h]
\includegraphics[scale=0.4]{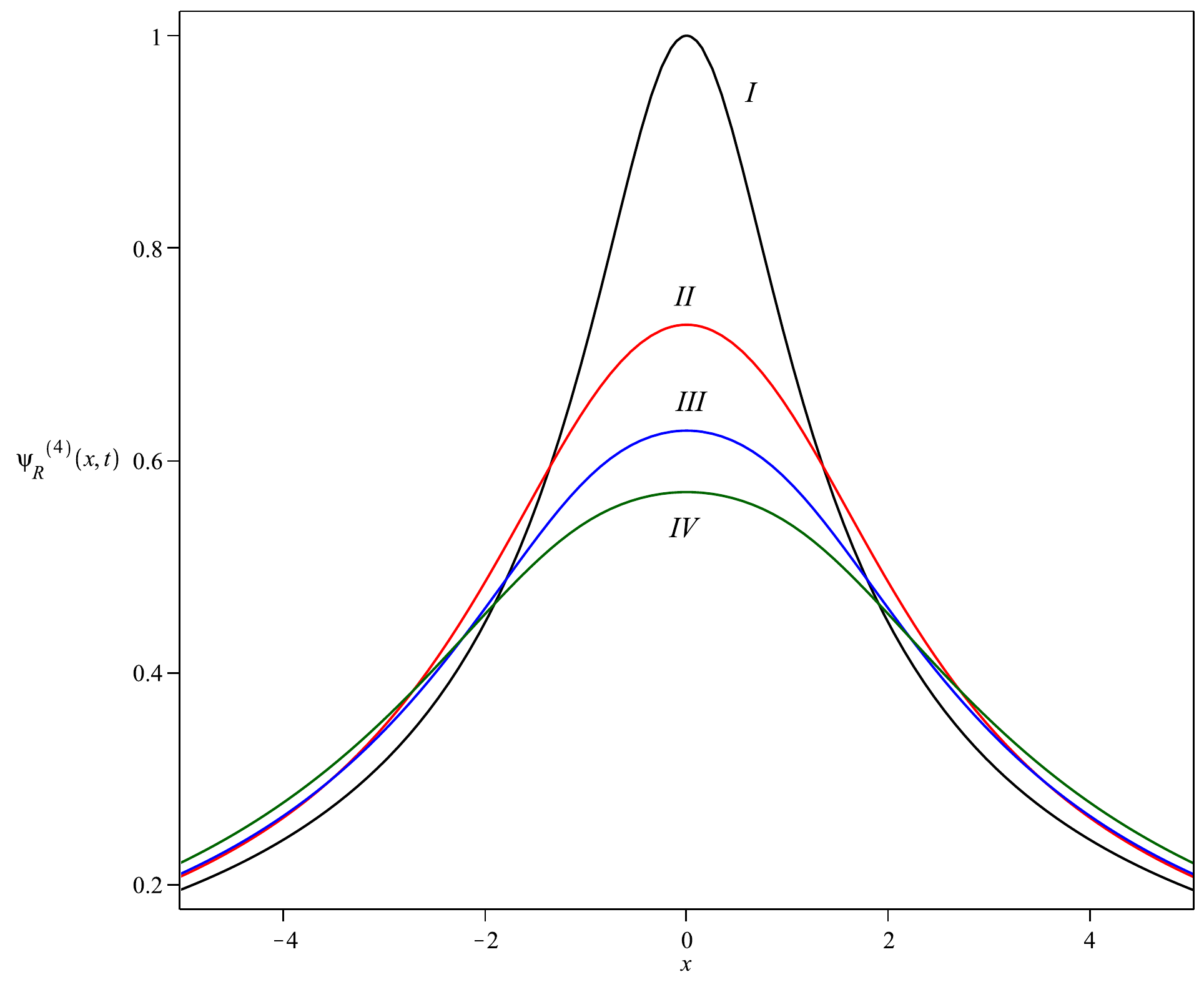}
\caption{\label{fig4} (Color online) Plot of $\psi_{\rm R}^{(4)}(x, t)$ given by Eq. \eqref{30/07-19} for $t=0$ (black line, I), $t=2$ (red line, II), $t=4$ (blue line, III), and $t=6$ (green line, IV).}
\end{figure}

The treatment of the next set of solutions evolving from the (non-normalizable) i.c. $f^{(4)}(x) = (1+x^{2})^{-1/2}$, $|x| < \infty$, is in a sense similar to the preceding case. The key tool to evaluate Eq. \eqref{30/07-7} here is Eq. (1.1.2.42) of \cite{YuABrychkov} which reads: $\partial_{x}^{2n}[(1+x^{2})^{-1/2}] = (2n)! (1+x^{2})^{-(n+1/2)}P_{2n}(\ulamek{x}{\sqrt{1+x^{2}}})$, with $P_{k}(z)$ the Legendre polynomials \cite{NIST}, $|z|<1$. Accordingly, the solution is given in integral form as
\begin{align}\label{30/07-17}
\psi^{(4)}(x, t) & = \frac{\E^{t}}{\sqrt{1+x^{2}}}\! \int_{0}^{\infty}\!\! g_{1/2}(y) \E^{-y t^{2}}\!\! \left[\sum_{n=0}^{\infty}\! \left(\!\frac{y t^{2}}{1+x^{2}}\!\right)^{n}\right.\nonumber \\
& \times \left.\Gamma(n+\ulamek{1}{2}) P_{2n}(\ulamek{x}{\sqrt{1+x^{2}}})\right]\D y,
\end{align}
whose inner summation $r^{(4)}(\tilde{t}, \tilde{x})\equiv\sum_{n=0}^{\infty} \tilde{t}^{n}\Gamma(n+\ulamek{1}{2})\times\\ P_{2n}(\tilde{x})$ can be performed with the use of Eq. (4) from \S 108 of \cite{Rain} and Eq. (5.12.1.4) of \cite{APPrudnikov-v2}, with the result:
\begin{equation}\label{30/07-18}
r^{(4)}(\tilde{t}, \tilde{x}) = \int_{0}^{\infty} \frac{\exp\left(\frac{-u^{2}-\tilde{t} u^{2} (1-\tilde{x}^{2})}{1 + \tilde{t} u}\right)}{\sqrt{u(1 + \tilde{t} u)}} \D u,
\end{equation}
$|\tilde{x}| \leq 1$. Defining now $R^{(4)}(t, x, y) \equiv r^{(4)}(\ulamek{4 y t^{2}}{1+x^{2}}, \ulamek{x}{\sqrt{1+x^{2}}})$ we end up with
\begin{equation}\label{30/07-19}
\psi^{(4)}(x, t) = \int_{0}^{\infty}\!\! g_{1/2}(y) \E^{-y t^{2}} \frac{\E^{t} R^{(4)}(t, x, y)}{\sqrt{\pi(1 + x^{2})}} \D y.
\end{equation}
Several examples of $\psi^{(4)}(x, t)$ are displayed on the Fig.~\ref{fig4}.

After having worked out the R evolution from the usual set of i.c. $f^{(j)}(x)$, $j=1, \ldots, 4$ we shall apply our formalism to a less common i.c., namely $f^{(5)}(x) = g_{1/2}(x)$, $x \geq 0$, with $\int_{0}^{\infty} x^{2} f^{(5)}(x)dx$ diverging. (This is fully justified as $g_{1/2}^{(k)}(x)\vert_{x=0} = 0$ for all $k = 0, 1, \ldots$.) The application of Eq. \eqref{30/07-9} immediately yields: 
\begin{align}\label{30/07-20}
& \psi^{(5)}_{\rm R}(x, t) = \E^{t} \int_{0}^{\infty} g_{1/2}(u) \E^{-u t^{2}} \left\{\frac{1}{2t\sqrt{\pi u}} \right. \nonumber\\
& \qquad \times \left.\int_{-\infty}^{\infty}\exp\left[-\frac{(x-\xi)^{2}}{4ut^{2}}\right]g_{1/2}(\xi) \D\xi\right\}\D u \\
& \qquad = \frac{t\E^{t}}{2\pi^{3/2}} \int_{0}^{\infty} \frac{K_{1}(\sqrt{t^{2} + (x-\xi)^{2}})}{\xi^{3/2}\sqrt{t^{2} + (x-\xi)^{2}}} \E^{-\frac{1}{4\xi}} \D\xi, \label{30/07-21}
\end{align}
where in Eq. \eqref{30/07-20} we switched the order of integrations over $u$ and $\xi$, and then employed Sommerfeld's representation of modified Bessel functions $K_{\nu}(z)$, s. formula (8.432.6) of \cite{RyGr}. Observe that Eq. \eqref{30/07-21} is fully consistent with the path-integral evaluation of the corresponding quantity in \cite{Kl}. In Fig. \ref{fig5} we display $\psi^{(5)}_{\rm R}(x, t)$ and $\psi_{\rm NR}^{(5)}(x, t)$ for several values of $t$. Observe that $f^{(5)}(x)$, initially confined to $0 \leq x < \infty$, spreads out to negative $x$ for \textit{any} arbitrarily small $t$. Here again, the R evolution is ``slower'' as the NR one, as seen by comparing the R curves with the $\psi_{\rm NR}^{(5)}(x, t)$ solutions for the \textit{same} values of $t$, obtained via the GW transform of Eq. \eqref{30/08-1} with~$f^{(5)}(x)$. \\
\begin{figure}[!h]
\includegraphics[scale=0.46]{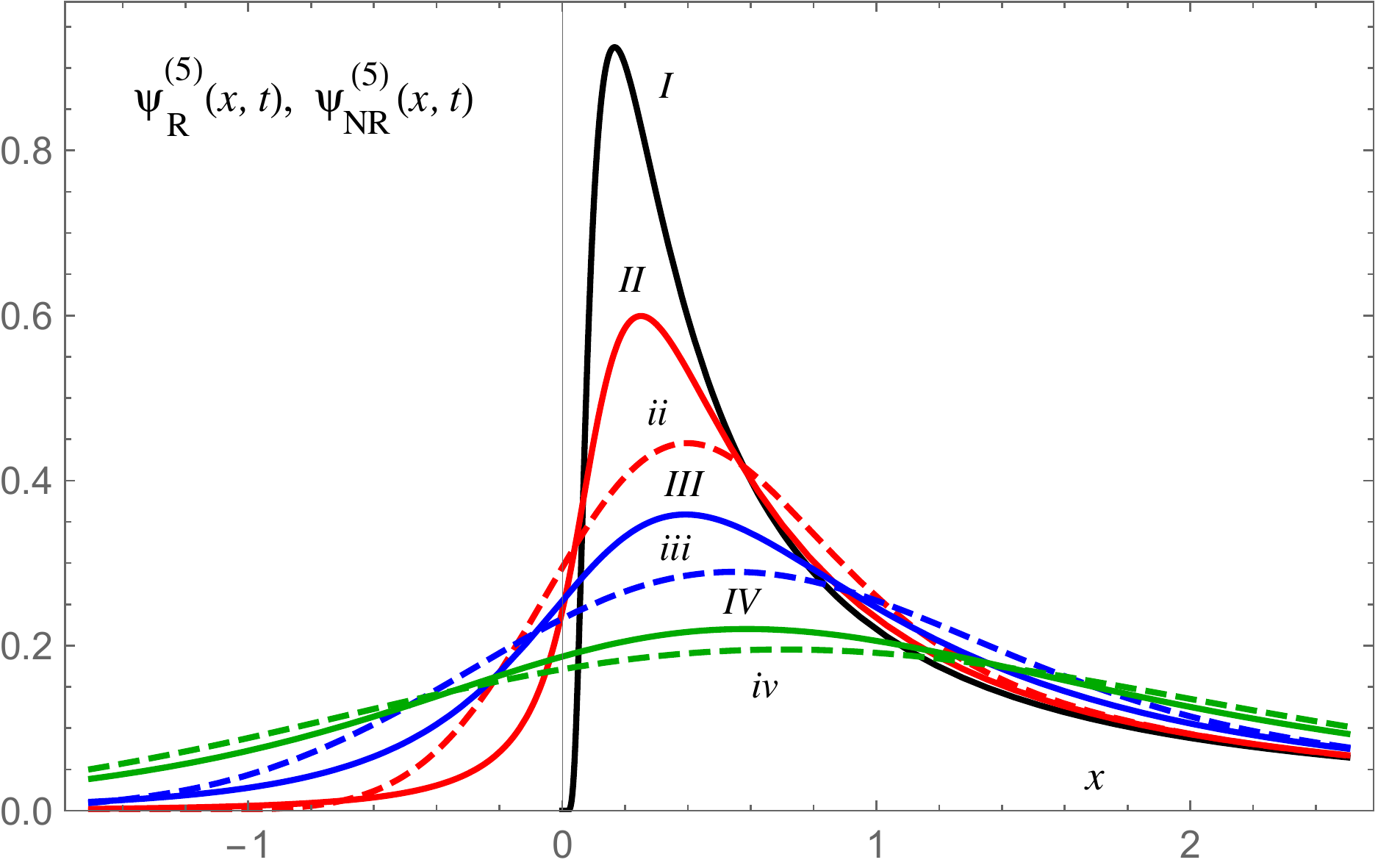}
\caption{\label{fig5} (Color online) Plot of $\psi_{\rm R}^{(5)}(x, t)$ given by Eq. \eqref{30/07-21} (solid curves) and $\psi_{\rm NR}^{(5)}(x, t)$ given by Eq. \eqref{30/08-1} (dashed curves) with $g_{1/2}(x)$ as the i.c. (black solid line, I) for $t=1/8$ (red line, II and ii), $t=1/2$ (blue line, III and iii), and $t=3/2$ (green line, IV and iv).}
\end{figure}

The problem of a consistent extension of the conventional diffusion to the relativistic situation has been extensively discussed for a long time. For a detailed chronology of these efforts consult \cite{JoPr}. The main reason was to properly describe the heat waves with finite speeds \cite{Catt, JoPr, IgnOS}. Amongst other subjects to which the relativistic-type diffusion equations would be relevant we mention the Landau theory of the second sound in Helium II \cite{Lan, Pesh} and in more general phonon gases \cite{Chest, Krum}, as well as the thermoelasticity \cite{IgnOS, Pov, Agar}. The initial hope that the Cattaneo-type equations would be appropriate to treat this range of problems was dissipated once it turned out that this approach is plagued with inherent inconsistencies \cite{CJ}. The current prevailing opinion is that the problem is still open \cite{DH, CJ} and therefore the study of other options is evidently legitimate.

The salient feature of our exact analysis based on Eq. \eqref{30/07-2} is the observation that the R diffusion propagates more slowly that the NR one while still retaining its ``normal" character. It will be intriguing to verify if this feature persists in more realistic settings. The natural extension of our formalism would involve considerations in higher dimensions and the explicit inclusion of the potential energy, which constitutes a true challenge. 


\section*{Acknowledgments}
K. G., A. H. and K. A. P. were supported by the PAN-CNRS program for French-Polish collaboration. Moreover, K. G. thanks for support from MNiSW (Warsaw, Poland), "Iuventus Plus 2015-2016", program no IP2014 013073.


\end{document}